\documentclass[prd,aps,showpacs,superscriptaddress,nofootinbib,amsmath,amssymb]{revtex4}

 \usepackage{graphicx}% Include figure files
\usepackage{dcolumn}% Align table columns on decimal point
\usepackage{bm}% bold math
\usepackage{lipsum}
\begin{document}

%\begin{frontmatter}

\title{Parity violating asymmetry with nuclear medium effects in deep inelastic $\vec e$ scattering}
\author{H. Haider}
\author{M. Sajjad Athar}
\email{sajathar@gmail.com}
\author{S. K. Singh}
\affiliation{Department of Physics, Aligarh Muslim University, Aligarh - 202 002, India}
\author{I. \surname{Ruiz Simo}}
\affiliation{Departamento de F\'{\i}sica At\'omica, Molecular y Nuclear,
and Instituto de F\'{\i}sica Te\'orica y Computacional Carlos I,
Universidad de Granada, Granada 18071, Spain}
\begin{abstract}
Recently parity violating asymmetry($A_{PV}$) in the Deep Inelastic Scattering(DIS) of polarised electrons($\vec e$) from deuterons has been measured at JLab 
and there exists future plans to measure this asymmetry from nuclear targets. In view of this we study nuclear medium effects in $A_{PV}$
in the DIS of $\vec e$ from some nuclear targets like
$^{12}C$, $^{56}Fe$ and $^{208}Pb$. The effects of Fermi motion, binding energy and nucleon correlations are taken into account through the 
 nucleon spectral function calculated in a local density approximation for nuclei. The pion and rho cloud contributions have also been taken into account. 
 This model has been earlier used to study nuclear medium effects in the electromagnetic and weak interaction induced processes.
 The results are presented and discussed in view of the future JLab experiments. 
\end{abstract}
\maketitle
\section{Introduction} 
 There is enough evidence from the experimental and theoretical studies of Deep Inelastic Scattering(DIS) of charged leptons and 
neutrinos from nuclear targets, to show that the quark Parton Distribution Function(PDF) for nucleons bound in nuclei are different from quark PDF in 
free nucleon. This is because the nucleons present in the nucleus have constraints due to Fermi motion, binding energy, 
nucleon off mass shell and nucleon correlation effects which are important in the high region of Bjorken variable x dominated by valence quarks. 
In the small x region of Bjorken variable dominated by sea quarks, non-nucleonic degrees of freedom like pions and/or quark clusters and nuclear shadowing
play an important role. 
 However, the models which rely on pion excess to explain the DIS of leptons on nuclei
 are unable to explain the observed low x behavior seen in Drell-Yan processes where no significant enhancement has been experimentally seen 
 for the nuclear targets. 
 A description of nuclear effects which can consistently explain the observed effects in DIS and DY processes has been lacking. 
 
 Parity Violation in Electron Scattering (PVES) is another process which has been used to probe the quark structure of nucleons. The
 first observation of parity violating asymmetries in the DIS of polarised electrons with the deuteron target at SLAC~\cite{Prescott:1979dh} confirmed the Standard Model of Electroweak 
 Interactions in electron sector. 
 The most recent experiments at JLab performed with the polarised electrons on the deuteron targets in the DIS~\cite{nature} and the resonance~\cite{wang,G0:2011aa}
 regions have measured parity violating asymmetries with a very high precision and have determined the weak electron quark couplings, $g_{A,V}^{e,q} (=2C_{1u}-C_{1d})$ 
 and $g_{V,A}^{e,q}(=2C_{2u}-C_{2d})$. 
 Specifically, the high precision achieved in these experiments has made it possible, for the first time, to determine $g_{V,A}^{e,q}$~\cite{nature}. This 
 also verifies the phenomenon of quark-hadron duality in the weak sector of inclusive electron scattering~\cite{wang}.
 Earlier experiments in low and medium energy region performed 
 at MAINZ~\cite{mainz,Maas:2004dh,Maas:2004ta}, MIT-BATES~\cite{Beise:2004py,mitbates}, JLab~\cite{Armstrong:2005hs,Acha:2006my,Androic:2009aa} and SLAC~\cite{slac}
 have made significant contributions to the study of various aspects of quark structure of nucleon. For example, weak charge of the proton, vector and axial vector strangeness form factors of the nucleon, neutron densities of nuclei and neutral current transition form factors of 
 $N-\Delta$ transition. These have been summarized in many review papers, for example see Refs.~\cite{GonzalezJimenez:2011fq,Armstrong2012,Kumar2013}.

 In the high energy region specially in the DIS region, PVES experiments provide direct access to the study of weak electron and quark couplings to the Z boson, 
 (anti)quark parton distributions in nucleons and their modifications in nuclei. With the precision achieved in the present experiments, it may be possible to 
 measure parity violating asymmetry with high precision in future experiments planned with 12GeV upgrade at JLab~\cite{jlabupdate}. A comparison with 
 the state of the art
 theoretical calculations would be able to explore the physics beyond the Standard Model(BSM) in electroweak processes. 
 With the aim of studying the parity violating effects
 with high precision, experiments with Hydrogen, Deuterium, and other nuclear targets like Fe, Au and Pb are planned~\cite{pvdis_proposal}. Theoretically, 
 following the first calculations of the parity violating asymmetry in the DIS region done by Cahn and Gilman~\cite{cahn} in the Bjorken limit, various corrections to the asymmetry
 arising due to higher twist effect, finite $Q^2$ evolution of (anti)quark PDF, target mass correction, charge symmetry violation and nuclear medium effects have been done by many 
 authors~\cite{Brady:2011uy,thomas,PV_res,Matsui:2005ns,Gorchtein:2011mz,Mantry,Hobbs}. In future experiments to be performed with 
 nuclear targets to study parity violating asymmetries, it will be important to understand the nuclear medium effects as emphasized by Cloet et al.~\cite{thomas}. 
   In view of these theoretical and experimental developments, we have studied in this paper nuclear medium effects in parity violating asymmetry, arising due to binding energy, 
 Fermi motion, nucleon correlation, mesonic degrees
of freedom of nuclei and target mass correction. We also discuss the effect due to non-isoscalarity of nucleus on PV asymmetry in nuclei like $^{56}Fe$ and $^{208}Pb$
 for which nuclear medium effects have been recently found to be important in a mean field approximation using NJL Lagrangian~\cite{nambu,njl1,njl2}.
 
 When electron scatters from a bound nucleon in a nucleus and has Fermi momentum described by a momentum distribution, the Bjorken variable for the target parton
 acquires a Fermi momentum dependence. The quark and antiquark PDFs should therefore be convoluted with the momentum distribution of the nucleon which takes into 
 account various nuclear medium effects in order to calculate the structure functions, entering in the expression for PV asymmetry. Moreover, there may be additional contribution due 
 to mesonic degrees of freedom in nuclei which may contribute to these structure functions as they do in EMC effect of charged leptons and neutrinos. 
 We calculate these nuclear medium effects using a model which has been earlier applied with fair degree of success in describing the EMC effect 
 and cross section data on $\nu$-scattering on $^{12}C$, $^{56}Fe$ and $^{208}Pb$~\cite{athar,sajjadplb,prc84,prc85}.
 We have also applied this model to study the effects of non-isoscalarity in Paschos-Wolfenstein relation in the nuclear medium~\cite{prc87} as well as in studying nuclear medium effects in the 
 Drell-Yan process~\cite{DY}.
 
 In section-2, we briefly describe the essential expression for the asymmetry in terms of structure functions calculated in nuclear medium along with the contribution
 from mesonic degrees of freedom. In section-3,  we present the numerical results and summarize our findings in section-4.

\section{Parity Violating Asymmetry}
\subsection{Formalism}
The parity violating asymmetry $A^{\rm PV}$ in the scattering of polarised electron from nucleon/nucleus is defined as
 \begin{equation*}
A^{\rm PV}=\frac{\sigma^R-\sigma^L}{\sigma^R+\sigma^L},
\label{eq:APV1}
\end{equation*}
where $\sigma^{R(L)}= \frac{d^2\sigma^{R(L)}}{d\Omega' dE'}$ denotes the scattering cross section for the right(left) handed polarised electron.
The asymmetry arises due to the interference between photon($\gamma$) and Z-boson($Z^0$) exchange amplitudes~\cite{Mantry}.

 The differential cross-section for electron-proton scattering takes the general form as~\cite{Mantry}
\begin{equation}\label{diff-cross}
\frac{d^2\sigma}{d\Omega' dE'} = \frac{\alpha^2}{Q^4}\frac{E'}{E} \Big ( L_{\mu \nu}^\gamma W_\gamma^{\mu \nu} - \frac{G_FQ^2}{4\sqrt{2} \pi \alpha} L_{\mu \nu}^{\gamma Z}W_{\gamma Z}^{\mu \nu}   \Big ),
\end{equation}
where $E$ and $E'$ denote  the energies of the incoming  and outgoing electrons respectively in the lab frame.  $Q^2=-q^2=-(\ell -\ell ')^2$ is the 
four momentum transfer square. $l_\mu $ and $l_\mu'$ denote the four-momenta of the incoming and outgoing electrons respectively.  The leptonic tensors in Eq.~(\ref{diff-cross}) are given by
\begin{eqnarray}
L_{\mu \nu}^\gamma &=&  2 (l_\mu l_\nu^\prime + l^\prime_\mu l_\nu - l \cdot l^\prime  g_{\mu \nu}- i \lambda \epsilon_{\mu \nu \alpha \beta} l^\alpha l^{\prime\beta} ), \nonumber \\
L_{\mu \nu}^{\gamma Z}&=& (g_V^e + \lambda g_A^e)L_{\mu \nu}^\gamma,
\end{eqnarray}
where $\lambda$ denotes the sign of the initial electron helicity with $\lambda=1$(right handed) and $\lambda=-1$(left handed). 

 The hadronic tensors $W_{\mu \nu}^{\gamma}$ and $W_{\mu \nu}^{\gamma Z}$ are parameterized in terms of dimensionless structure functions as:
\begin{eqnarray}\label{g-sfn}
W_{\mu \nu}^{\gamma} &=& \left ( -g_{\mu \nu} + \frac{q_\mu q_\nu}{q^2}\right)
\frac{F_1^\gamma}{M} + \left(P_\mu - \frac{P.q}{q^2} q_\mu \right)
\left (P_\nu - \frac{P.q}{q^2}q_\nu \right)\frac{F_2^\gamma}{ M P.q},
\end{eqnarray}
\begin{eqnarray}\label{gZ-sfn}
W_{\mu \nu}^{\gamma Z} &=& \left( -g_{\mu \nu} + \frac{q_\mu q_\nu}{q^2}\right)
\frac{F_1^{\gamma Z}}{M} + \left(P_\mu - \frac{P.q}{q^2} q_\mu \right) 
\left (P_\nu - \frac{P. q}{q^2}q_\nu \right)\frac{F_2^{\gamma Z}}{ M P.q}+ 
\frac{i \epsilon_{\mu \nu \alpha \beta} P^\alpha q^\beta}{2 M P \cdot q} F_3^{\gamma Z}.~~~~~ 
\end{eqnarray}
In the Bjorken limit ($Q^2, \nu \to \infty$, $x$ fixed),
the interference structure functions $F_1^{\gamma Z}$ and
$F_2^{\gamma Z}$ are related by the Callan-Gross relation,
$F_2^{\gamma Z} = 2x F_1^{\gamma Z}$, similar to the electromagnetic
$F_{1,2}^\gamma$ structure functions $F_2^\gamma(x) = 2x F_1^\gamma(x)$. 

$F_1^\gamma$ is given in terms of nucleon PDFs as~\cite{Hobbs}:
\begin{equation}
\label{eq:Fg}
F_1^\gamma(x) = \frac{1}{2} \sum_q e_q^2\ (q(x) + \bar{q}(x)); ~~F_2^\gamma(x) = x \sum_q e_q^2\ (q(x) + \bar{q}(x)) , 
\end{equation}
for the pure electromagnetic case, while

\begin{subequations}
\label{eq:FgZ}
\begin{align}
F_1^{\gamma Z}(x) &= \sum_q e_q\ g_V^q\ (q(x) + \bar{q}(x)); ~~F_2^{\gamma Z}(x) = 2x\sum_q e_q\ g_V^q\ (q(x) + \bar{q}(x)), 	\\		
F_3^{\gamma Z}(x) &= 2 \sum_q e_q\ g_A^q\ (q(x) - \bar{q}(x))\ ,
\end{align}
\end{subequations}
are the structure functions occurring in the weak-electromagnetic interference term. For the numerical calculations, we have used parton distribution 
functions of CTEQ6.6~\cite{cteq}. The vector couplings for the u and d quarks are given respectively by $g^u_V = -1/2 + (4/3)sin^2\theta_W$ and
 $g^d_V =  1/2 - (2/3)sin^2\theta_W$, while the quark axial-vector couplings are $g^u_A =  1/2$ and $g^d_A = -1/2$, respectively. 

In terms of these structure functions $F_i^\gamma(i=1-2)$ and $F_i^{\gamma Z}(i=1-3)$, the PVDIS asymmetry(APV) can be written as:
\begin{equation}\label{eq:APV}
A^{\rm PV}
=  -\left( {G_F Q^2 \over 4 \sqrt{2} \pi \alpha} \right)
  { g^e_A
    \left( 2xy F_1^{\gamma Z} - 2 [1 - 1/y + xM/E] F_2^{\gamma Z} \right)
  + g^e_V\
    x (2-y) F_3^{\gamma Z}
  \over
    2xy F_1^\gamma - 2 [1 - 1/y + xM/E] F_2^\gamma
  }\ .
\end{equation}
where $y=\nu/E$ is the lepton fractional energy loss. 

The PV asymmetry in Eq.~(\ref{eq:APV}) may also be written as:
\begin{equation}\label{apvy}
A^{\rm PV}
=  \left( {G_F Q^2 \over 4 \sqrt{2} \pi \alpha} \right)
    \left( Y_1\ a_2\ +\ Y_3\ a_3 \right)\ ,
\end{equation}
where $a_2$ is given by:
\begin{subequations}
\label{eq:a13}
\begin{equation}\label{eq:a1}
a_2(x) = -2g_A^e\frac{F_{2}^{\gamma Z}(x)}{F_{2}^\gamma(x)}
\end{equation}
while $a_3$ is given by:
\begin{equation}\label{eq:a33}
a_3(x) = -2\,x g_V^e \frac{F_{3}^{\gamma Z}(x)}{F_{2}^{\gamma}(x)}
\end{equation}
\end{subequations}
where $g^e_V( = -\frac{1}{2} + 2sin^2\theta_W$) is the vector and $g^e_A( = -\frac{1}{2})$ is the axial-vector couplings of the charged lepton.

$a_2$ and $a_3$ terms may also be written in terms of quark PDFs as
\begin{eqnarray}\label{a_2}
a_2(x) = \frac{2\sum_q e_q g_V^q q_A^+(x)}{\sum_q e_q^2 q_A^+(x)}.
\end{eqnarray}
\begin{eqnarray}\label{a_3}
a_3(x) = -4 g_V^e \frac{\sum_q e_q g_A^q q_A^-(x)}{\sum_q e_q^2 q_A^+(x)},
\end{eqnarray}
where $q_A^+(x_A)=q_A(x_A)~+~\bar q_A(x_A)$, $q_A^-(x_A)=q_A(x_A)~-~\bar q_A(x_A)$ and $q=u,d,s,c$. Here the subscript A stands for a nucleus.

If we expand $a_2$ about $u_A \simeq d_A$ limit and $s_A^+, c_A^+  << u_A^+ + d_A^+$, then one may write~\cite{thomas}
\begin{eqnarray}
a_2(x) \simeq \frac{9}{5} - 4\sin^2\theta_W
- \frac{12}{25}\,
\frac{u_A^+(x) - d_A^+(x)-s_A^+(x)+c_A^+(x)}{u_A^+(x) + d_A^+(x)}.
\label{eq:a2mod}
\end{eqnarray}

Similarly for $a_3(x_A)$
\begin{eqnarray}
a_3(x) &\simeq& \frac{9}{5}\left[
 1 - 4\sin^2\theta_W \right] \times \nonumber\\ 
 &&\left[\frac{u_A^-(x) + d_A^-(x)}{u_A^+(x) + d_A^+(x)} + \frac{1}{3}\frac{u_A^-(x) - d_A^-(x)}{u_A^+(x) + d_A^+(x)} - \frac{3}{5}\frac{u_A^+(x) - d_A^+(x)}{u_A^+(x) + d_A^+(x)} 
 - \frac{2}{5}\frac{s_A^+(x)}{u_A^+(x) + d_A^+(x)} - \frac{8}{5}\frac{c_A^+(x)}{u_A^+(x) + d_A^+(x)} \right]~~
\label{eq:a3mod}
\end{eqnarray}

At finite $Q^2$, $R^{\gamma (\gamma Z)}$ are given in terms of the ratio of the longitudinal to transverse
virtual photon cross sections that may be written as:
\begin{equation}
R^{\gamma (\gamma Z)}\
\equiv\ \frac{\sigma_L^{\gamma (\gamma Z)}}{\sigma_T^{\gamma (\gamma Z)}}\ 
=\ r^2 \frac{F_2^{\gamma (\gamma Z)}}{2x F_1^{\gamma (\gamma Z)}} - 1\ ,
\end{equation}
for both the electromagnetic ($\gamma$) and interference ($\gamma Z$)
contributions, with
\begin{equation}
r^2 = 1 + {Q^2 \over \nu^2} = 1 + {4 M^2 x^2 \over Q^2}\ .
\end{equation}
In Eq.\ref{apvy}, $Y_1$ and $Y_3$ are given by ~\cite{Mantry}:
\begin{subequations}
\label{eq:Y}
\begin{align}
\label{eq:Y1}
Y_1
&= \frac{ 1+(1-y)^2-y^2 (1-r^2/(1+R^{\gamma Z})) - 2xyM/E }
	{ 1+(1-y)^2-y^2 (1-r^2/(1+R^\gamma)) - 2xyM/E } 
   \left( \frac{1+R^{\gamma Z}}{1+R^\gamma} \right)\ ,	\\
\label{eq:Y3}
Y_3
&= \frac{ 1-(1-y)^2 }
	{ 1+(1-y)^2-y^2 (1-r^2/(1+R^\gamma)) - 2xyM/E }
   \left( \frac{r^2}{1+R^\gamma} \right)\ .
\end{align}
\end{subequations}

In the Bjorken limit, $Y_1 = 1$, $Y_3 = \frac{1-(1-y)^2}{1+(1-y)^2}$ and the kinematical ratio $r^2 \to 1$. 
%***************************************************
\begin{figure}
\includegraphics[width=6cm,height=4cm]{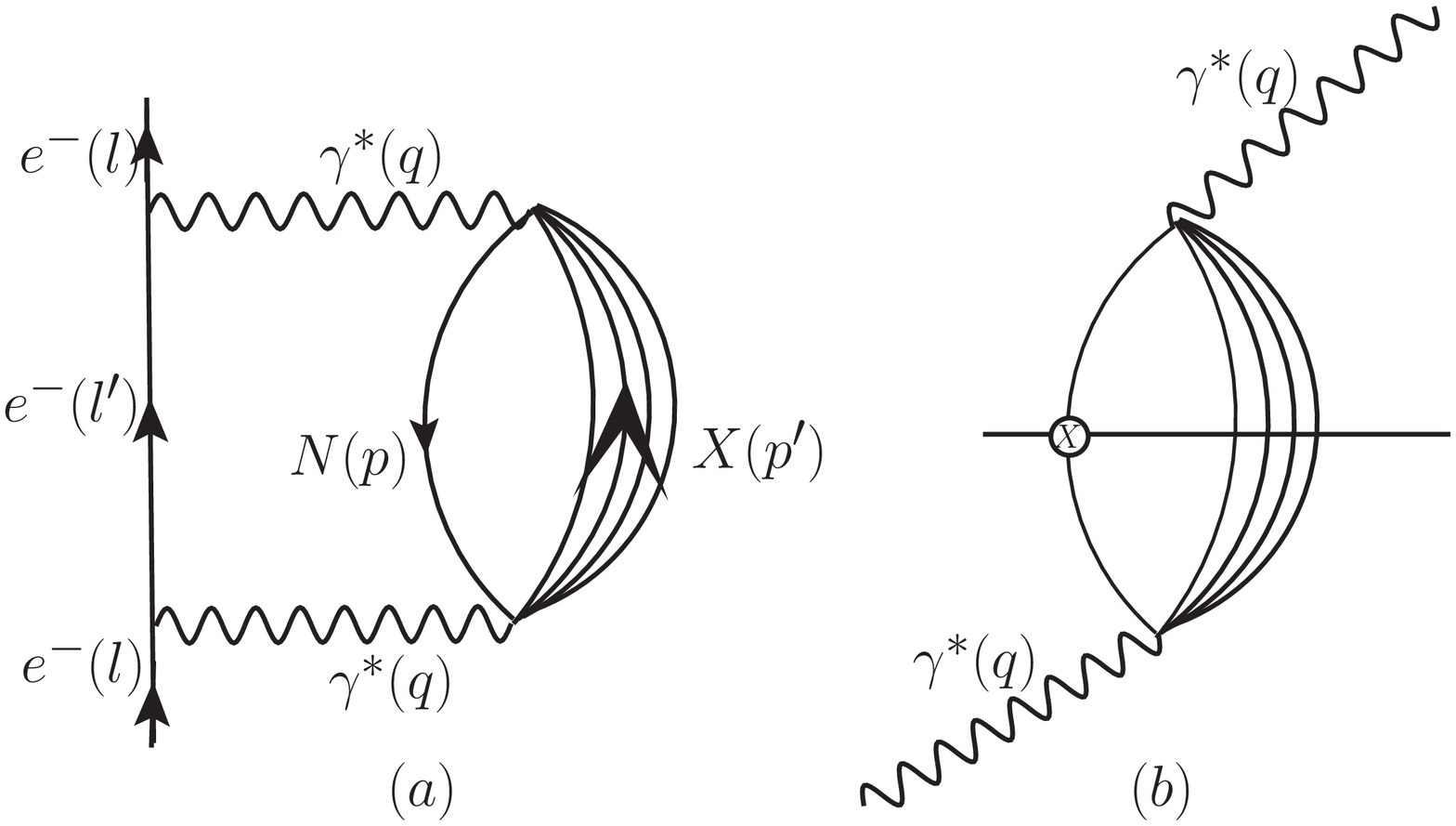}
\caption{(a) Electron self energy (b) Photon self energy. The imaginary part is calculated by cutting along the horizontal line and applying the Cutkosky rules while putting the particle on 
mass shell.}
\label{ese}
\end{figure}

\begin{figure}
\begin{center}
\includegraphics[width=6cm,height=4cm]{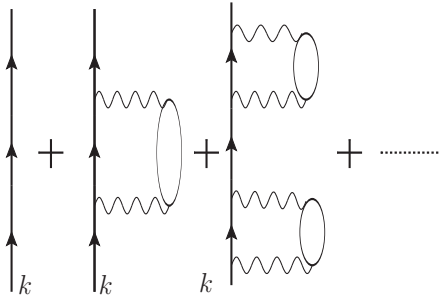}
\caption{Nucleon self-energy in the nuclear medium}
\label{3}
\end{center}
\end{figure}

\begin{figure}
\begin{center}
\includegraphics[height=12cm, width=12cm]{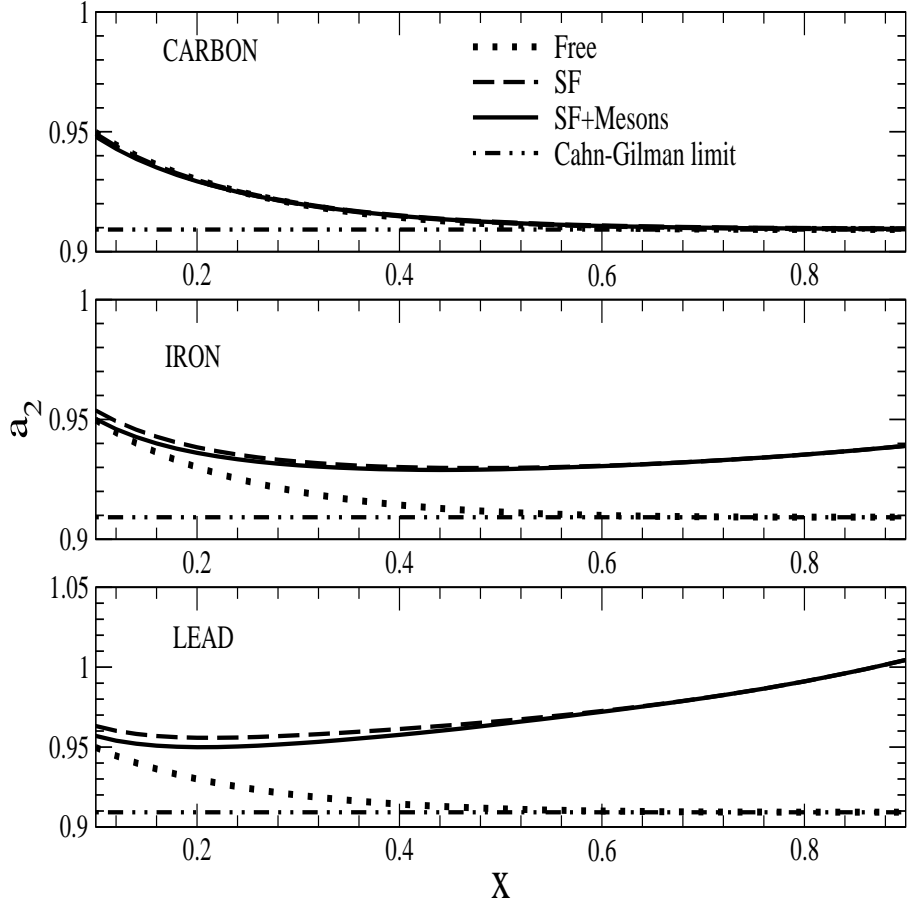}
\caption{$a_2(x)$ vs x in $^{12}C$(isoscalar), $^{56}Fe$(nonisoscalar) and $^{208}Pb$(nonisoscalar) nuclear targets at $Q^2=5 GeV^2$. 
The results for the free isoscalar nucleon case($a_2(x)=\frac{a_2^p(x)~+~a_2^n(x)}{2}$) have been shown by the dotted line(free), dashed-double dotted line
depicts the results in the Cahn-Gilman limit. The results with nuclear structure effects where we have taken into account the Fermi motion, Pauli blocking and
nucleon correlations have been shown by the dashed line(SF) and the solid line(SF+Mesons) is the result when the full prescription is used where the contribution 
of the meson cloud is also added.}
\label{fig1}
\end{center}
\end{figure}

\begin{figure}
\begin{center}
\includegraphics[height=9cm, width=14cm]{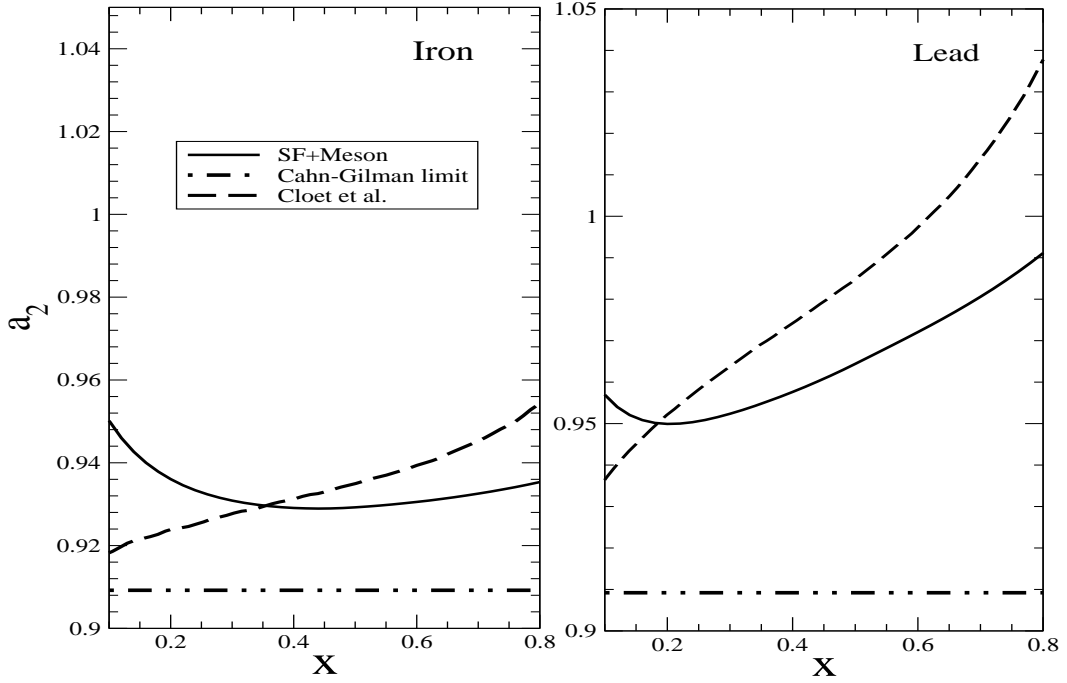}
\caption{$a_2(x)$ vs x at $Q^2=5 GeV^2$ for $^{56}Fe$(nonisoscalar) and
$^{208}Pb$(nonisoscalar) nuclear targets. The solid line is the result obtained using the 
full model(SF+Mesons), dashed line is the result of Cloet et al.~\cite{thomas} and the
 dashed-double dotted line is the result obtained in the Cahn-Gilman limit.}\label{fig2}
\end{center}
\end{figure}

If we neglect sea quark effects, and assume an isoscalar nucleus(N=Z) then in the Cahn-Gilman limit~\cite{cahn}, $a_2(x)$, $a_3(x)$ and $A_{PV}(x)$ are written as: 
\begin{eqnarray}\label{apv}
 a_2&=&  \frac{9}{5} - 4 sin^2\theta_W \nonumber\\
 a_3&=& \frac{9}{5} \left[1 - 4 sin^2\theta_W~\right] \nonumber\\
 A_{PV}&=&\frac{G_F Q^2}{4\sqrt{2}\pi\alpha}\frac{9}{5}\left[1 - \frac{20}{9} sin^2\theta_W~+~(1-4 sin^2\theta_W)\frac{1-(1-y)^2}{1+(1-y)^2}~\right]
\end{eqnarray}

\begin{figure}
\begin{center}
\includegraphics[height=12cm, width=10cm]{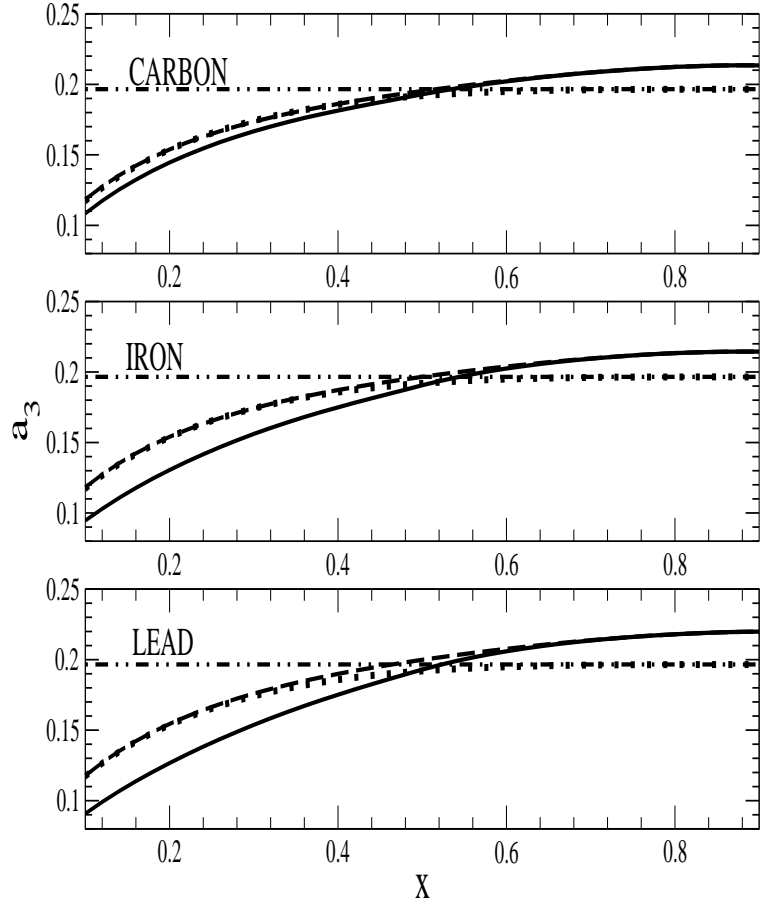}
\caption{$a_3(x)$ vs x at $Q^2=5 GeV^2$ for $^{12}C$(isoscalar), $^{56}Fe$(nonisoscalar) and $^{208}Pb$(nonisoscalar) nuclear targets. Lines and points have the same meaning as in 
Fig.\ref{fig1}.}
\label{fig3}
\end{center}
\end{figure}

\begin{figure}
\begin{center}
\includegraphics[height=6cm, width=10cm]{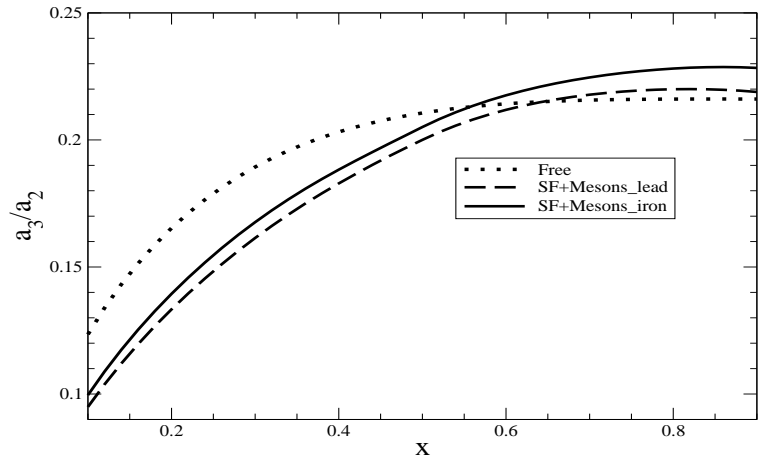}
\caption{$\frac{a_3}{a_2}$ vs x at $Q^2=5 GeV^2$ for $^{56}Fe$(solid line) and $^{208}Pb$(dashed line) nuclei using full model(Spectral Function+Meson cloud contribution).
Dotted line is the result for the free isoscalar nucleon(free) case.}
\label{fig4}
\end{center}
\end{figure}

\subsection{Target Mass Correction}
At low $Q^2$ and for high values of x the scattering kinematics are modified by the nucleon mass and therefore the nucleon structure functions $F_{2,3}(x,Q^2)$ are modified. 
In the present work, the target mass correction(TMC) has been taken from the works of Schienbein et al.~\cite{schienbein}, and the modified structure functions are given by
\begin{equation}\label{f2TMC}
 F_{2}^{TMC}(x,Q^2)\simeq\frac{x^2}{\xi^2\,\gamma^3} F_{2}(\xi)\left[ 1+\frac{6\, \mu\, x\, \xi}{\gamma}(1-\xi)^2\right],
\end{equation}
and
\begin{equation}\label{f3TMC}
F_{3}^{{\rm TMC}}(x,Q^{2}) \simeq \frac{x}{\xi \gamma^{2}} F_{3}(\xi)
\bigg[1-\frac{\mu x \xi}{\gamma} (1-\xi) \ln \xi \bigg]\, .
\end{equation}
where $\mu=\frac{M^2}{Q^2}$, $ \gamma = \sqrt{ 1 + \frac{4 x^2 M^2}{ Q^2 } }$ and $\xi$ is the Nachtmann variable defined as $\xi = \frac{2 x}{1+\gamma}$.
We have used Eq.\ref{f2TMC} for $F_2^{\gamma,\gamma Z}$ and Eq.\ref{f3TMC} for $F_3^{\gamma Z}$ to incorporate target mass correction.
\subsection{Nuclear Medium Effects}\label{sec:NE}
\subsubsection{Nuclear Structure}
  When the reaction takes place on a 
nucleon target inside the nucleus, several nuclear effects like Fermi motion, binding energy, nucleon correlations, pion and rho meson cloud contributions, etc. must be taken into account. Presently we have
implemented Fermi motion, nucleon binding energy and nucleon correlations through the use of a nucleon spectral function. For this we have used local density approximation (LDA). This model has been successfully used earlier to describe the photon, 
lepton and neutrino induced reactions in the intermediate energy region for the nuclei of the present interest~\cite{athar,sajjadplb,prc84,prc85,Marco}. We are using a relativistic formalism for an interacting Fermi sea and the local density
approximation is used to translate results from nuclear matter to finite nuclei. The use of nucleon Green's 
function in terms of their spectral functions offers a way to account
for the Fermi motion, binding energy of the nucleon inside the nucleus and nucleon correlations. Therefore, we construct a relativistic nucleon spectral function and define everything
within a field theoretical formalism which uses the nucleon propagators 
written in terms of this spectral function.
The required nuclear information needed is contained in the nucleon spectral function. 

Like Eq.~\ref{diff-cross}, we write the differential cross-section for electron-nucleus scattering as
\begin{equation}\label{diff-cross-A}
\frac{d^2\sigma^A}{d\Omega' dE'} = \frac{\alpha^2}{Q^4}\frac{E'}{E} \Big ( L_{\mu \nu}^\gamma W_{\gamma,~A}^{\mu \nu} - \frac{G_FQ^2}{4\sqrt{2} \pi \alpha} L_{\mu \nu}^{\gamma Z}
W_{_{\gamma Z,~A}}^{\mu \nu} \Big ),
\end{equation}
where  $W_{\gamma,~A}^{\mu \nu}$ and $W_{_{\gamma Z,~A}}^{\mu \nu}$ now correspond to nuclear hadronic tensors.  These nuclear hadronic tensors are given in terms of dimensionless nuclear structure 
functions $F_i^A$, as
\begin{eqnarray}\label{g-sfnA}
W_{\mu \nu,~A}^{\gamma} &=& \left ( -g_{\mu \nu} + \frac{q_\mu q_\nu}{q^2}\right)
\frac{F_{1~A}^\gamma}{M_{_A}} + \left(P_{_A,\mu} - \frac{P_{_A}.q}{q^2} q_\mu \right)
\left (P_{_A,\nu} - \frac{P_{_A}.q}{q^2}q_\nu \right)\frac{F_{2~A}^\gamma}{ M_{_A} P_{_A}.q},
\end{eqnarray}
\begin{eqnarray}\label{gZ-sfnA}
W_{\mu \nu,~A}^{\gamma Z} &=& \left( -g_{\mu \nu} + \frac{q_\mu q_\nu}{q^2}\right)
\frac{F_{1~A}^{\gamma Z}}{M_{_A}} + \left(P_{_A,\mu} - \frac{P_{_A}.q}{q^2} q_\mu \right) 
\left (P_{_A,\nu} - \frac{P_{_A}. q}{q^2}q_\nu \right)\frac{F_{2~A}^{\gamma Z}}{ M_{_A} P_{_A}\cdot q}+ 
\frac{i \epsilon_{\mu \nu \alpha \beta} P_{_A}^\alpha q^\beta}{2 M_{_A} P_{_A} \cdot q} F_{3~A}^{\gamma Z}.~~~~~ 
\end{eqnarray}
In the above expression $P_A^\mu=(M_{_A}, {\bf 0})$ is the four momentum of the nucleus.

In our formalism the lepton nuclear cross sections are obtained in terms of lepton self energy $\Sigma$ in the nuclear medium. For example, if we consider electron-nucleon DIS, the 
electron self energy corresponding to the diagram shown in Fig.\ref{ese}a), is obtained by considering the cross section $d\sigma$ for an element of volume $dV$ in the nucleus as
\begin{equation}\label{defxsec}
d\sigma=\Gamma dt ds=\Gamma\frac{dt}{dl}ds dl=\Gamma \frac{1}{v}dV = \Gamma \frac {E_e}{\mid \vec k \mid}dV=\frac{-2m_e}{E_e(\vec k)} Im \Sigma (k)\frac{E_e(k)}{\mid \vec{k} \mid}dV,
\end{equation}
where $\Gamma$ is the electron width and $\Sigma$ is the electron self energy. Thus to get $d\sigma$, we are required to evaluate imaginary part of electron self energy i.e $Im \Sigma (k)$.

Following Feynman rules~\cite{Itzykson}, $\Sigma (k)$ for Fig.~\ref{ese}a) may be written as
\begin{eqnarray}
- i \Sigma (k) = \sum_{spin}\int \frac{d^4 q}{(2 \pi)^4} \; 
\bar{u_e} (\vec{k}) \; i e \gamma^{\mu} \; i \frac{\not \! k' + m}{k'^{2} -
m^2 + i \epsilon} \; i e \gamma^{\nu} u_e (\vec{k}) \; 
\frac{-i g_{\mu \rho}}{q^2} \; (- i) \; \Pi^{\rho \sigma} (q) \; 
\frac{-i g_{\sigma \nu}}{q^2}
\end{eqnarray}
where $\Pi^{\mu \nu} (q)$ is the photon self energy.

For an unpolarized electrons the above expression may be written as
\begin{equation}\label{imsigma}
\Sigma (k) = i e^2 \; \int \frac{d^4 q}{(2 \pi)^4} \;
\frac{1}{q^4} \;
\frac{1}{2m} \;
L_{\mu \nu} \; \frac{1}{k'^2 - m^2 + i \epsilon} \; \Pi^{\mu \nu} (q)
\end{equation}
where $L_{\mu\nu}$ is the Leptonic tensor. Im$\Sigma$ (k) is obtained by taking the imaginary part of the right hand side of the above 
equation which also contains photon self energy $\Pi^{\mu \nu} (q)$, the details of which may be found in Ref.~\cite{Marco}.

When Eq.\ref{imsigma} is used along with Eq.\ref{defxsec}, we obtain 
\begin{equation}\label{dsigma_3}
\frac {d\sigma^A}{d\Omega'dE'}=-\frac{\alpha}{q^4}\frac{k'}{k}\frac{1}{(2\pi)^2} L_{\mu\nu} \int  Im \Pi^{\mu\nu}d^{3}r
\end{equation}
It may be noted that when Eq.~\ref{diff-cross-A} and Eq.~\ref{dsigma_3} are compared we may express nuclear hadronic tensor $W_A^{\mu \nu}$ in terms of $Im \Pi^{\mu\nu}$. 

 The photon self energy $\Pi^{\mu \nu} (q)$ corresponding to Fig.~\ref{ese}b) is written in terms of nucleon propagator, meson propagator and the 
 hadronic current which contains nucleon structure functions. Using Feynman rules this is written as,
\begin{eqnarray}\label{photonse}
- i \Pi^{\mu \nu} (q)&=&( - ) \; \int \frac{d^4 p}{(2 \pi)^4} \; i G (p) \;
\Sigma_X \; \Sigma_{s_p, s_i} \Pi^N_{i = 1}
\int \frac{d^4 p'_i}{(2 \pi)^4}
\\ \nonumber
&&
\Pi_l i G_l (p'_l) \Pi_j \; i D_j (p'_j) 
( - i )^2 e^2 < X | J^{\mu} | H > < X | J^{\nu} | H >^*
(2 \pi)^4  \; \delta^4 (q + p - \Sigma^N_{i = 1} p'_i)
\end{eqnarray}
where $G_l$ is the nucleon propagator and $D_j$ is the meson propagator. These nucleons and mesons are moving in a nuclear medium where because of the medium effects the free propagators 
are replaced by the dressed propagators. This is done in the following ways.

The relativistic Dirac propagator $G_l (p'_l)$ for nucleon in an interacting Fermi sea is derived in terms of free nucleon Dirac propagator $G^{0}(p_{0},{{\bf p}})$ which is written as
\begin{equation}  \label{prop2}
G^{0}(p_{0},{{\bf p}}) =\frac{M}{E(p)}\left\{\frac{\sum_{r}u_{r}(p)\bar u_{r}(p)}{p^{0}-E({{\bf p}})+i\epsilon}+\frac{\sum_{r}v_{r}(-p)\bar v_{r}(-p)}{p^{0}+E({{\bf p}})-i\epsilon}\right\}
\end{equation}
where $u({\bf p})$ and $v({\bf p})$ are the Dirac spinors~\cite{Itzykson}.

The relativistic nucleon propagator in a non-interacting Fermi sea may be written as
\begin{eqnarray}  \label{prop4}
G^{0}(p_{0},{{\bf p}})&=&\frac{M}{E({{\bf p}})}\left\{\sum_{r}u_{r}(p)\bar u_{r}(p)
\left[\frac{1-n(p)}{p^{0}-E({{\bf p}})+i\epsilon}+\frac{n(p)}{p^{0}-E({{\bf p}})-i\epsilon}\right]+\frac{\sum_{r}v_{r}(-p)\bar v_{r}(-p)}{p^{0}+E({{\bf p}})-i\epsilon}\right\}
\end{eqnarray}
where $n({{\bf p}})$ is the occupation number of nucleons in the Fermi sea, $n({{\bf p}})$=1 for p$\le p_{F_N}$ while $n({{\bf p}})$
=0 for p$> p_{F_N}$. 

We shall retain only the positive energy contributions as the negative energy contributions are suppressed. In the interacting Fermi sea, 
 the relativistic nucleon propagator is written using Dyson series expansion (depicted in Fig.\ref{3}) in terms of nucleon self energy $\sum(p^0,p)$. 
 This perturbative expansion is summed in ladder approximation to give~\cite{Marco}:
\begin{eqnarray}
G(p)&=&\frac{M}{E({\bf p})}\sum_{r}u_{r}({\bf p})\bar u_{r}({\bf p})\frac{1}{p^{0}-E({\bf p})}+\frac{M}{E({\bf p})}\sum_{r}\frac{u_{r}({\bf p})\bar u_{r}({\bf p})}{p^{0}-E({\bf p})}\sum(p^{0},{\bf p})\frac{M}{E({\bf p})}\sum_{s}\frac{u_{s}({\bf p})\bar u_{s}({\bf p})}{p^{0}-E(p)}+.....\\ \nonumber
&&=\frac{M}{E({\bf p})}\sum_{r}\frac{u_{r}({\bf p})\bar u_{r}({\bf p})}{p^{0}-E({\bf p})-\bar u_{r}({\bf p})\sum^N(p^{0},{\bf p})u_{r}({\bf p})\frac{M}{{\bf E(p)}}}
\end{eqnarray}
The expression for the nucleon self energy in nuclear matter i.e. $\sum(p^0,p)$ is taken from Ref.~\cite{FernandezdeCordoba:1991wf}.
The relativistic nucleon propagator G(p) in a nuclear medium is expressed in terms of hole  $S_h (\omega, \bbox{p})$ and particle $S_p (\omega, \bbox{p})$ spectral functions as~\cite{Marco,FernandezdeCordoba:1991wf}:
\begin{eqnarray}\label{Gp}
G (p) =&& \frac{M}{E({\bf p})} 
\sum_r u_r ({\bf p}) \bar{u}_r({\bf p})
\left[\int^{\mu}_{- \infty} d \, \omega 
\frac{S_h (\omega, \bbox{p})}{p^0 - \omega - i \eta}
+ \int^{\infty}_{\mu} d \, \omega 
\frac{S_p (\omega, \bbox{p})}{p^0 - \omega + i \eta}\right]\,,
\end{eqnarray}
where $S_h (\omega, \bbox{p})$ and $S_p (\omega, \bbox{p})$ being the hole
and particle spectral functions respectively, which are given by~\cite{FernandezdeCordoba:1991wf}:
\begin{equation}\label{sh}
S_h (\omega, {\bf p})=\frac{1}{\pi}\frac{\frac{M}{E({\bf p})}Im\Sigma^N(p^0,p)}{(p^0-E({\bf p})-\frac{M}{E({\bf p})}Re\Sigma^N(p^0,p))^2
+ (\frac{M}{E({\bf p})}Im\Sigma^N(p^0,p))^2}
\end{equation}
for $p^0 \le \mu$
\begin{equation}
S_p (\omega, {\bf p})=-\frac{1}{\pi}\frac{\frac{M}{E({\bf p})}Im\Sigma^N(p^0,p)}{(p^0-E({\bf p})-\frac{M}{E({\bf p})}Re\Sigma^N(p^0,p))^2
+ (\frac{M}{E({\bf p})}Im\Sigma^N(p^0,p))^2}
\end{equation}
for $p^0 > \mu$.

$\mu$ is the chemical potential. This prescription has been earlier
 applied to study photo, charged lepton and neutrino induced processes in nuclei~\cite{Marco,sajjadplb,athar,prc84,prc85,prc87}.

 Before proceeding ahead, first we ensure that the spectral function which are presently being used is properly normalized. For this we follow the method of 
 Frankfurt and Strikman~\cite{Frankfurt} where the baryon number conservation is imposed. Furthermore, using this spectral function we also try to get binding 
 energy for a given nucleus to be very close to the experimental value. To apply these constrains we proceed in the following way. 
 
 The electromagnetic form factor at 
 q=0 is evaluated assuming baryons having unit charge which corresponds to
\begin{eqnarray}\label{NNmu}
\left<N|B^{\mu}|N\right>\equiv \bar u({{\bf p}}) \gamma^{\mu} u({{\bf p}})~=~B \frac{p^{\mu}}{M};~ B=1,~ p^{\mu}\equiv(E({{\bf p}}),{{\bf p}})
\end{eqnarray}
When the nucleons are in the nuclear medium it can shown that\cite{Marco,FernandezdeCordoba:1991wf} 
\begin{eqnarray}\label{AAmu}
\left<A|B^{\mu}|A\right>=- Lt_{_{\eta \rightarrow 0^+}} \int \frac{d^4p}{(2\pi)^4}V i Tr[G(p^0,{{\bf p}})\gamma^\mu]e^{ip^0\eta}.
\end{eqnarray}
where V is the volume of the normalization box and $exp({ip^0\eta})$, with $\eta$, is the convergence factor for loops 
appearing at equal times.

Using the expression given by Eq. \ref{Gp}, it can be seen that the convergence factor limits the contribution to the hole spectral 
function and one gets
\begin{eqnarray} \label{ABmu}
\left<A|B^{\mu}|A\right>&=&V\int\frac{d^{3}p}{(2\pi)^3}\frac{M}{E({{\bf p}})}{\vec Tr} \left[\sum_{r}u_{r}({{\bf p}})\bar u_{r}({{\bf p}})\gamma^{\mu}\right]\int_{-\infty}^{\mu}S_{h}(\omega,p)d\omega \nonumber\\
&&=V\int\frac{d^{3}p}{(2\pi)^3}\frac{M}{E({{\bf p}})}{\vec Tr} \left[\frac{(\not p+M)_{on shell}}{2M}\gamma^{\mu}\right]\int_{-\infty}^{\mu}S_{h}(\omega,p)d\omega \nonumber\\
&&=2V\int\frac{d^{3}p}{(2\pi)^3}\frac{M}{E({{\bf p}})}\frac{p^{\mu}_{on shell}}{M}\int_{-\infty}^{\mu}S_{h}(\omega,p)d\omega\equiv B\frac{P_A^\mu}{M_A} 
\end{eqnarray}
It is to be noted that in the last step we have imposed that this matrix element gives the right current with $B$ baryons, in analogy to 
the expression given by Eq \ref{NNmu}. For a nucleus at rest the above expression implies
\begin{eqnarray} \label{norm1}
2V\int\frac{d^{3}p}{(2\pi)^{3}}\int_{-\infty}^{\mu}S_{h}(\omega,p) d\omega= B~=1
\end{eqnarray}

In the local density approximation, the spectral
functions of protons and neutrons are the function of local Fermi momentum.
The equivalent normalization to Eq.\ref{norm1} is written as
\begin{eqnarray} \label{norm2}
2\int\frac{d^{3}p}{(2\pi)^{3}}\int_{-\infty}^{\mu}S_{h}(\omega,p,k_{F_{p,n}}({\vec r})) d\omega= \rho_{p,n}({\vec r})
\end{eqnarray}
where factor 2 in the above expression is due to two possible projections of spin $\frac{1}{2}$ particle. $k_{F_{p(n)}}$ is the Fermi momentum of proton(neutron) inside the nucleus which is 
expressed in terms of proton(neutron) densities given by $k_{F_{p(n)}}({\vec
r})= \left[ 3\pi^{2} \rho_{p(n)}({\vec r}) \right]^{1/3}$. These nucleon densities are in turn related with the nuclear densities 
$\rho(r)$(like $\rho_p(r)=\frac{Z}{A}\rho(r)$ and $\rho_n(r)=\frac{A-Z}{A}\rho(r))$, the parameters of which are determined 
from electron scattering experiments. In the present calculation we have used harmonic
 oscillator density for $^{12}$C nucleus, two parameter Fermi  density for $^{56}$Fe and $^{208}Pb$ nuclei, and the density parameters are taken from Refs.~\cite{DeJager:1987qc}-\cite{GarciaRecio}.
 
For a symmetric/isoscalar nuclear matter of density $\rho({\vec r})$, there is a unique Fermi momentum given by $k_{F}({\vec
r})= \left[ 3\pi^{2} \rho({\vec r})/2\right]^{1/3}$ for which we obtain
\begin{eqnarray} \label{norm3}
4\int\frac{d^{3}p}{(2\pi)^{3}}\int_{-\infty}^{\mu}S_{h}(\omega,p,k_{F}({\vec r})) d\omega= \rho({\vec r})
\end{eqnarray}
 
 Eq.~\ref{norm2} leads to the normalization condition individually satisfied by proton and neutron as
\begin{eqnarray}\label{norm4}
2 \int d^3 r \;  \int \frac{d^3 p}{(2 \pi)^3} \int^{\mu}_{- \infty} \; S_h (\omega, \vec{p}, \rho_p(r)) \; d \omega &=& Z\,,\nonumber\\
2 \int d^3 r \;  \int \frac{d^3 p}{(2 \pi)^3} \int^{\mu}_{- \infty} \; S_h (\omega, \vec{p}, \rho_n(r)) \; d \omega &=& A-Z\,,~~\rm {and}\nonumber\\
4 \int d^3 r \;  \int \frac{d^3 p}{(2 \pi)^3} \int^{\mu}_{- \infty} \; S_h (\omega, \vec{p}, \rho(r)) \; d \omega &=& A~~~\rm{for~an ~isoscalar ~nucleus}.
\end{eqnarray}

Also we calculate the average kinetic and total nucleon energy given by: 
\begin{eqnarray}
<T>= \frac{4}{A} \int d^3 r \;  \int \frac{d^3 p}{(2 \pi)^3} (E({\bf p})-M) 
\int^{\mu}_{- \infty} \; S_h (p^0, {\bf p}, \rho(r)) 
\; d p^0\,,
\end{eqnarray}
\begin{eqnarray}
<E>= \frac{4}{A} \int d^3 r \;  \int \frac{d^3 p}{(2 \pi)^3}  
\int^{\mu}_{- \infty} \; S_h (p^0, {\bf p}, \rho(r)) 
\; p^0 d p^0\,,
\end{eqnarray}
and the binding energy per nucleon(B.E./A) given by~\cite{Marco}:
\begin{equation}
|E_A|=-\frac{1}{2}(<E-M>+\frac{A-2}{A-1}<T>)
\end{equation}
Here we have ensured that the kinetic energy $<T>$ and the total energy $<E>$ for the nucleon inside the nucleus are retrieved. 
We have tabulated in Table~\ref{table1}, the kinetic energy per nucleon ($< T >/A$) and the binding energies per nucleon (B.E./A) for Carbon, Iron and Lead nuclei.
\begin{table}
\begin{center}

\begin{tabular}{|c|c|c|c|}\hline\hline
Nucleus  & $< T >/A$ (MeV) & B.E./A (MeV)\\
 \hline\hline
$^{12}$C & 20.2 & 7.6  \\\hline
$^{56}$Fe & 30.0 & 8.8  \\\hline
$^{208}$Pb & 32.7 & 7.8  \\\hline\hline
\end{tabular}
\end{center}
\caption{ Kinetic energy per nucleon ($< T >/A$) and binding energy per nucleon (B.E./A) for $^{12}$C, $^{56}$Fe and $^{208}$Pb.}
\label{table1}
\end{table}

\begin{figure}
\begin{center}
\includegraphics[height=12cm, width=10cm]{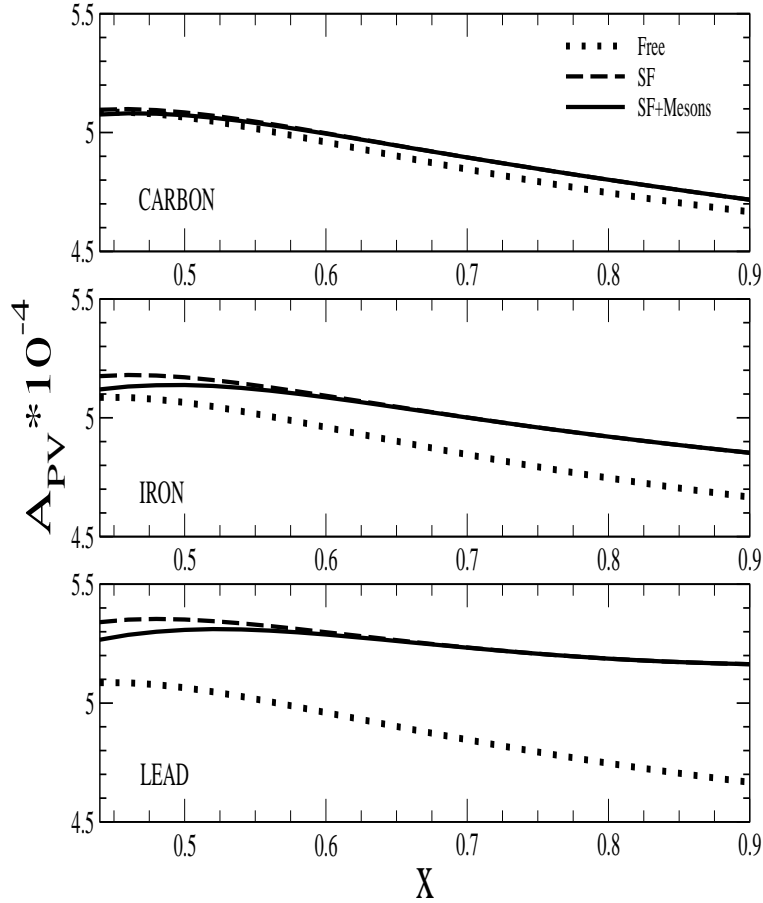}
\caption{Parity Violating Asymmetry $A_{PV}$ vs x, at $Q^2=5 GeV^2$ and electron beam energy E=6.06GeV in $^{12}C$(isoscalar), 
$^{56}Fe$(nonisoscalar) and $^{208}Pb$(nonisoscalar) nuclear targets(Lines and points have the same meaning as in Fig.\ref{fig1}).}\label{fig5}
\end{center}
\end{figure}
Once we make it sure that the spectral function for a given nucleus is properly normalized to give the baryon number and the correct binding energy, we write the 
nuclear hadronic tensor $W^{\alpha \beta}_{A~~\gamma}$
 and $W^{\alpha \beta}_{A~~W}$ in terms of the corresponding nucleonic tensors  $W^{\alpha \beta}_{\gamma}$ and $W^{\alpha \beta}_{W}$ as 
 
 \begin{equation}	\label{conv_WA}
W^{\alpha \beta}_{A~~\gamma,~\gamma Z} = 4 \int \, d^3 r \, \int \frac{d^3 p}{(2 \pi)^3} \, 
\frac{M}{E ({\bf p})} \, \int^{\mu}_{- \infty} d p^0 S_h (p^0, {\bf p}, \rho(r))
W^{\alpha \beta}_{~~\gamma,~\gamma Z} (p, q), \,.
\end{equation}
where the nucleon structure function $W^{\alpha \beta} (p, q)$ is given by Eq.~\ref{g-sfn} for the $\gamma$ exchange and by Eq.~\ref{gZ-sfn} for the $\gamma Z$ term.

Using Eqs.~\ref{g-sfn}, \ref{g-sfnA} and \ref{conv_WA}, for electromagnetic case and Eqs.~\ref{gZ-sfn}, \ref{gZ-sfnA} and \ref{conv_WA}, for 
electroweak interference case, we obtain~\cite{athar, Marco}
\begin{eqnarray}\label{F2A}
F_{2~A}^{\gamma(\gamma Z)}(x_A,Q^2)=4\int d^3r\int \frac{d^3p}{(2\pi)^3}\int_{-\infty}^\mu d\omega\;S_h(\omega,\mathbf{p},\rho(\mathbf{r}))\frac{\left(1-\gamma\frac{p_z}{M}\right)}{\gamma^2}
\left(\gamma'^2+\frac{6x_N^2(\mathbf{p}^2-p^2_z)}{Q^2}\right){F_{2~N}^{\gamma(\gamma Z)}}(x_N,Q^2)
\end{eqnarray}
with $p^0=M+\omega$, $\gamma'^2=1+4x_N^2p^2/Q^2$, $x_N=Q^2/(2p\cdot q)$ and $x_A=\frac{x}{A}=\frac{1}{A}\frac{Q^2}{2Mq_0}$.

Similarly $F_{3~A}^{\gamma Z}$ nuclear structure function is derived to be
\begin{eqnarray}\label{F3A}
F_{3~A}^{\gamma Z}(x_A,Q^2)&=&4\int d^3r\int \frac{d^3p}{(2\pi)^3}\frac{M}{E(\mathbf{p})}\int_{-\infty}^\mu d\omega\;S_h(\omega,\mathbf{p},\rho(\mathbf{r}))\left(\frac{p_0\gamma-p_z}{(p_0-p_z\gamma)\gamma}\right)
{F_{3~N}^{\gamma Z}}(x_N,Q^2)~~~~~~~
\end{eqnarray}

The expression for $F_{2~A}^{\gamma(\gamma Z)}$ given in Eq.\ref{F2A} has been derived assuming the validity of Callan-Gross relation 
between nuclear structure functions $F_{1A}$ and $F_{2A}$. The violation of Callan-Gross relation may arise through the contribution of higher twist(HT) effects due to quark-quark and quark-gluon 
 interactions in Quantum Chromodynamics(QCD) as well as due to mesonic contributions in the nuclear medium.  It has been shown that higher twist(HT) effects do not 
 contribute to the violation of Callan-Gross relation up to twist 4-level in QCD~\cite{Mantry}. The violation due to mesonic and other nuclear effects are 
 shown to be important only in the region of low x and low $Q^2$~\cite{Castorina:2002zf}. This region is not relevant for the present interest of parity 
 violating experiments from nuclear targets at JLab which are planned in the region of higher x and low $Q^2$~\cite{jlabupdate}. 
 
 However, the topic of violations of Callan-Gross relation and other sum rules like Gottfried sum rule, Adler sum rule, etc. in nuclei is of current interest 
 in itself and ongoing experiments at JLab~\cite{phdthesis,jlab} to measure $F_{1},~F_{2}$ and $F_{L}$ on nuclear targets will provide important 
 information on this subject~\cite{Boer:2011fh}. 
 
  Furthermore, the nuclear modification of $F_{3A}$ structure function is shown to be consistent to GLS sum rule which is in agreement 
  with the present experimental results~\cite{Kim:1998kia}  and other theoretical 
 results~\cite{athar}.
 
\subsubsection{$\pi$ and $\rho$ mesons contribution to the nuclear structure function}\label{Pion_Contribution}
The attractive interactions of nucleons inside the nucleus enhances the meson cloud and this increases the probability for an incident photon or Z boson to interact
with a pion or a rho meson instead of a nucleon. The pion and rho meson cloud contributions to the $F_2$ structure function have been implemented following the many body field theoretical approach
of Ref.~\cite{Marco}. In the case of $F_3^{\gamma Z}$ structure function there is no contribution from pion and rho meson clouds as $F_3^{\gamma Z}$ only gets contribution from valence quark distributions.
 The expression of pion structure function $F_{2,\pi}^A (x)$ and rho structure function $F_{2,\rho}^A (x)$ in the nucleus is obtained through same formalism as done for nuclear structure function by replacing the bound nucleon 
propagator by pion or rho meson propagator.

The pion and rho meson cloud contributions to the $F_2$ structure function have been implemented following the many body field theoretical approach described in Ref.~\cite{Marco}. 
The pion structure function $F_{2 , \pi}^A (x_A)$ is written as:
\begin{equation}  \label{F2pion}
F_{2~\pi~~A}^{\gamma(\gamma Z)}(x_A,Q^2) = - 6 \int  d^3 r  \int  \frac{d^4 p}{(2 \pi)^4} \; 
\theta (p^0) \; \delta I m D (p) \; 
\; \frac{x}{x_\pi} \; 2 M \; F_{2~\pi}^{\gamma(\gamma Z)} (x_\pi) \; \theta (x_\pi - x) \; 
\theta (1 - x_\pi) 
\end{equation}
where $D (p)$ the pion propagator in the medium given in terms of the pion self energy $\Pi_{\pi}$:
\begin{equation}
D (p) = [ {p^0}^2 - \vec{p}\,^{2} - m^2_{\pi} - \Pi_{\pi} (p^0, p) ]^{- 1}\,,
\end{equation}
where
\begin{equation}\label{pionSelfenergy}
\Pi_\pi=\frac{f^2/m_\pi^2 F^2(p)\vec{p}\,^{2}\Pi^*}{1-f^2/m_\pi^2 V'_L\Pi^*}\,.
\end{equation}
Here, $F(p)=(\Lambda^2-m_\pi^2)/(\Lambda^2+\vec{p}\,^{2})$ is the $\pi NN$ form factor and  $\Lambda$=1 GeV~\cite{athar}, $f=1.01$, $V'_L$ is
the longitudinal part of the spin-isospin interaction and $\Pi^*$ is the irreducible pion self energy that contains the contribution of particle - hole and delta - hole excitations.
In Eq.(\ref{F2pion}), $\delta Im D(p)$ is given by  

\begin{equation}
\delta I m D (p) \equiv I m D (p) - \rho \;
\frac{\partial Im D (p)}{\partial \rho} \left|_{\rho = 0} \right.
\end{equation}
and
\begin{equation} 
\frac{x}{x_{\pi}} = \frac{- p^0 + p^z}{M},~{\rm where}~x_\pi=-\frac{Q^2}{2p \cdot q}
\end{equation}

Following the same notation as in Ref.~\cite{Gluck:1991ey},  the pion structure function at LO can be written in terms of pionic PDFs as
\begin{eqnarray}\label{pion_structure_function}
F_{2\pi}^{\gamma}(x_\pi)&=&\sum_q e_q^{2} x_\pi\left[q(x_\pi)+\bar{q}(x_\pi)\right]\nonumber\\
F_{2\pi}^{\gamma~Z}(x_\pi)&=&\sum_q e_q x_\pi g_v^q \left[q(x_\pi)+\bar{q}(x_\pi)\right]
\end{eqnarray}
where  $q(x_\pi)$ and $\bar{q}(x_\pi)$ are the quark and antiquark parton distribution functions  in the pion.

Similarly, the contribution of the $\rho$-meson cloud to the structure function  is written as~\cite{Marco}

\begin{equation} \label{F2rho}
F_{2~\rho~~A}^{\gamma(\gamma Z)}(x_A,Q^2) = - 12 \int d^3 r \int \frac{d^4 p}{(2 \pi)^4}
\theta (p^0) \delta Im D_{\rho} (p) \frac{x}{x_{\rho}} \, 2 M
F_{2, \rho}^{\gamma(\gamma Z)} (x_{\rho}) \theta (x_{\rho} - x) \theta (1 - x_{\rho})
\end{equation}

\noindent
where $D_{\rho} (p)$ is the $\rho$-meson propagator and $F_{2 \rho}
(x_{\rho})$ is the $\rho$-meson structure function, which we have taken equal to the pion structure function $F_{2\pi}$ using the valence and sea pionic PDFs from reference \cite{Gluck:1991ey}. 
$\Lambda_\rho$ in $\rho NN$ form factor $F(p)=(\Lambda_\rho^2-m_\rho^2)/(\Lambda_\rho^2+\vec{p}\,^{2})$ has also 
been taken as 1 GeV.

It may be pointed out that our model also fulfills the momentum
sum rule as expressed in Eq. 88 of Kulagin and Petti~\cite{Kulagin1989}  (including also the $\rho$ meson).
The procedure is straight forward by using Eqs. 36-37 from Ref.~\cite{Kulagin1989} and expression given in Eqs.\ref{F2pion} and \ref{F2rho}.
The details of the prescription is given in Ref.~\cite{Kulagin1989}. 
The pion $<y>_\pi$ and nucleon $<y>_N$ fractions of the light cone momentum are related by
\begin{equation}
 <y>_\pi+ <y>_N = \frac{M_A}{A M},
\end{equation}
where $M_A$ is the nucleus mass. The nucleon quantities can be easily obtained from the spectral function.
 Our results are like this: For iron $< y >_n= 0.967$, $\pi$+$\rho$ should account for 0.024.

\section{Results and Discussion}
We have used Eq.\ref{F2A} for $F_{2~A}^{\gamma}$ and $F_{2~A}^{\gamma Z}$ and Eq.\ref{F3A} for $F_{3~A}^{\gamma Z}$, to evaluate 
nuclear structure functions. The mesonic contribution due to pion and rho mesons has been evaluated using Eq.\ref{F2pion} for $F_{2, \pi}^A (x)$ and Eq.\ref{F2rho} 
for $F_{2, \rho}^A (x)$. 
The (anti)quark PDF parameterizations for nucleon  as determined by the CTEQ collaboration~\cite{cteq} 
and the (anti)quark parameterizations for  pion(rho) mesons as given by Gluck et al.~\cite{Gluck:1991ey} have been used in the calculations.
The Target Mass Correction(TMC) has been incorporated using Eqs.18 and 19 in the appropriate structure functions and all the results presented here are with TMC. 
The asymmetry  $A_{\rm PV}$ is calculated using $a_2(x)$ and $a_3(x)$ from Eq.\ref{eq:a1} and Eq.\ref{eq:a33} respectively.
For numerical calculations, we have taken $sin^2\theta_W=0.2227$~\cite{thomas,zeller} and the results are presented for $^{12}C$, $^{56}Fe$ and $^{208}Pb$ nuclei.
Eq.~\ref{apv} has been used to present the results in the Cahn-Gilman limit for $a_2$ and $a_3$.

In Fig.\ref{fig1}, we show the results for $a_2(x)$, the term containing the contribution of quark vector coupling for 
$^{12}C$,  $^{56}Fe$ and $^{208}Pb$. The results for the free isoscalar nucleon case(shown by the dotted line) have been obtained by defining $a_2(x)=\frac{a_2^p(x)~+~a_2^n(x)}{2}$. 
 We find that
 
 (i) For an isoscalar nuclear target like $^{12}C$(top panel), there is an enhancement in $a_2(x)$ over the Cahn-Gilman limit at lower values of x(x$<$0.4). 
 This enhancement is obtained
 in the free isoscalar nucleon case also(dotted curve) when $a_2(x)$ is calculated using Eq.~\ref{eq:Fg} for $F_{2}^\gamma(x)$ and Eq.~\ref{eq:FgZ} for $F_2^{\gamma~Z}(x)$
 in Eq.~\ref{eq:a1}. This implies that the enhancement is mainly due to the sea quark content of the nucleon which are neglected in deriving the Cahn-Gilman limit.
 There is almost no change in $a_2(x)$ due to nuclear structure and mesonic effects (compare dashed and solid curves). In view of this there is almost no 
 dependence of change in $a_2(x)$ with the nucleon number A in the isoscalar limit $N=Z=\frac{A}{2}$. This has been numerically found to be less than $0.1\%$
 for the case of $^{24}Mg$ and $^{40}Ca$.

 (ii) For nonisoscalar nuclei like $^{56}Fe$(middle panel) and $^{208}Pb$(lower panel),  nuclear structure effects lead to an enhancement over the
 Cahn-Gilman limit in the entire region of x. This is over and above the enhancement due to sea quark of nucleons in the low x($<~0.4$) region and 
 is the main effect at large x($x~>~0.4$). This enhancement increases with the 
 nonisoscalarity and is smaller at lower values of x and becomes larger at high x. For example, 
 in the case of $^{56}Fe$ it is 1.5$\%$ at x=0.3 and becomes 3$\%$ at x=0.8 while in the case of $^{208}Pb$  it is 4$\%$ at x=0.3, which becomes 9$\%$ at x=0.8.
  
The inclusion of mesonic contribution leads to a very small decrease($<1\%$) only at smaller values of x $<~0.3$ for $^{56}Fe$ and x~$<~0.45$ for $^{208}Pb$. This is the region 
where mesonic effects are expected to be important but they seem to cancel out when the ratio $F_{2A}^{\gamma}$ to $F_{2A}^{\gamma Z}$ is taken.

In Fig.\ref{fig2}, we present our results for $^{56}Fe$ and $^{208}Pb$ in the full model (spectral function + meson cloud contribution) and compared them with the 
results of Cloet et al.~\cite{thomas} obtained in a valence quark model. The results in the Cahn-Gilman limit are also given for reference. We find that, our results are qualitatively similar 
to the results of Cloet et al.~\cite{thomas} in the region of higher x(x $>$ 0.3), dominated by the valence quarks. In this region of x, both the models predict an enhancement in the value of $a_2(x)$ over the Cahn-Gilman limit. This enhancement increases with the 
nonisoscalarity (N-Z) of nuclei as seen in the results presented for $^{56}Fe$ and $^{208}Pb$. However, 
quantitatively we find a smaller enhancement as compared to the results obtained by Cloet et al.~\cite{thomas}. This is because in the present model the 
enhancement is mainly due to the nuclear medium effects of Fermi motion, Pauli blocking and nucleon correlations included through nucleon spectral function as well 
as the mesonic contributions(pion and rho) in nuclei. On the other hand in the valence quark model of Cloet et al.~\cite{thomas}
 there is an initial enhancement present for the nucleons due to isospin dependence at quark level which gets further enhanced due to 
 nuclear medium effects treated through interaction of quarks with isospin dependent scalar and vector fields in NJL model~\cite{njl1,njl2}.
 
In Fig.\ref{fig3}, we have shown for the first time, the results for $a_3(x)$, the term containing the contribution of quark axial-vector couplings in the nuclear medium for 
 $^{12}C$(top panel), $^{56}Fe$(middle panel) and $^{208}Pb$(lower panel) and compared our results with the 
Cahn-Gilman limit of $a_3(x)$. For this, we have used Eq.~\ref{eq:Fg} for $F_{2}^\gamma(x)$ and Eq.~\ref{eq:FgZ} for $F_3^{\gamma Z}(x)$  
in Eq.~\ref{eq:a33} to get $a_3(x)$. We find that the nuclear medium effects lead to a suppression of $a_3(x)$ in the region of low 
x($x~<~0.5$) and an enhancement in the region of high $x$($x~>~0.5$).
The appreciable suppression seen in the low x region is also present in the isoscalar nucleon case 
$\left[a_3(x)=\frac{a_3^p(x)~+~a_3^n(x)}{2}\right]$(dotted line) shows that it is mainly due to nucleon sea quarks. In the high $x$ region, the enhancement is mainly due to nuclear medium effects.
The inclusion of mesonic contributions leads to a further decrease in $a_3(x)$ at low $x$ and almost no change 
in high $x$ region. 
  The relative changes in $a_3(x)$ due to nuclear medium effects (nuclear structure and mesonic) increases with nonisoscalarity. 
 For example, there is a further suppression of 7-8$\%$ over the free nucleon value at $x=0.2$ in $^{12}C$ which becomes 18-20$\%$ in
 $^{56}Fe$ and $^{208}Pb$. In the  high $x$ region, say $x=0.5$, the enhancement is $<~1\%$ for $^{12}C$ which becomes 2-3$\%$ for
 $^{56}Fe$ and $^{208}Pb$. This is because in the intermediate region of $x$($0.2~<~x~<~0.6$), $F_2^\gamma$ increases due to mesonic contribution while
 $F_3^\gamma$ has no mesonic contribution. In the region of 
high x, where sea quarks are not expected to play important role it is expected that  $a_3(x)$ would approach Cahn-Gilman limit.
 It is indeed so but since medium effects modify $F_{2A}^{\gamma, \gamma Z}$ and $F_{3A}^{ \gamma Z}$ differently due to nuclear structure effects 
 (see Eqs.\ref{F2A} and \ref{F3A}), a slightly higher value than the 
 Cahn-Gilman limit is obtained. We note that the nuclear medium effects lead to a suppression in the region of $x~<~0.5$ which is due to mesonic effects and increase 
 in the region of $x~>~0.5$ due to nuclear structure effects.

In Fig.\ref{fig4}, we show the ratio for $\frac{a_3(x)}{a_2(x)}$ vs x for $^{56}Fe$ and $^{208}Pb$ with nuclear structure and mesonic effects. 
The relative contribution of $a_3(x)$ to $a_2(x)$ is about $22\%$ at $x\approx 0.8$, which is approximately the same as in the free nucleon case and reduces
 to $12\%$ at $x \approx 0.2$ as compared to $15\%$ in the case of free nucleon. Thus in the region of present experimental interest (x$>$0.5), 
 the relative contribution 
 of $a_3(x)$ as compared to $a_2(x)$ is not affected much (about 2$\%$) due to the nuclear medium effects.
  
In Fig.\ref{fig5}, we show the results for $A_{PV}$ vs x using Eq.\ref{apvy} for $^{12}C$(top panel), $^{56}Fe$(middle panel) and $^{208}Pb$(lower panel) 
nuclear targets.
 We find the effect of nonisoscalarity as the largest source of nuclear medium effects, which
 enhances the asymmetry over the whole range of x. This enhancement increases with the excess of neutrons. For example, at $x=0.8$, the nuclear medium 
 effects give an increase of $3\%$
 in the case of $^{56}Fe$, while it becomes $8\%$ in the case of $^{208}Pb$. We have also studied the effect of modification in the lepton's energy 
 and momentum due to the presence of the Coulomb field of heavy nucleus like $^{56}Fe$ and $^{208}Pb$, using effective momentum approximation~\cite{Solvignon:2009it, Gueye:1999mm}. In this 
 approximation energy of incoming and outgoing electron which is in the few GeV energy region is modified by the average nuclear Coulomb 
 potential $V_0$. This potential $V_0$ is numerically of the order of a few tens of MeV in nuclei like  $^{56}Fe$ and $^{208}Pb$ and leads
 to a very small effect($<~1\%$) on $a_2$, $a_3$ and asymmetry $A_{PV}$.
 
 We have also included the contribution of heavy flavors and performed these calculations for the 3-quark flavors (up, down and strange) as well as with 4-quark flavors (up, down, strange and charm) and find the results to be within 1$\%$.
 The above results are presented at $Q^2=5GeV^2$. However, we have also obtained $A_{PV}$ at lower values of  $Q^2$ and found it to be almost independent of x.  
  Furthermore, we have found that the effect of TMC(not shown here) is to  increase the asymmetry by about 1-1.5$\%$ in the region of $0.3<x<0.8$.
\section{Summary and Conclusions}
In this paper, we have studied the nuclear medium effects on parity violating asymmetry $A_{PV}$ in the scattering of polarised electron from the nuclear targets like
 $^{12}C$, $^{56}Fe$ and $^{208}Pb$. Besides presenting numerical results for the asymmetry, we have also presented the numerical results for the terms $a_2(x)$ and $a_3(x)$, which determine
  the contributions of the quark vector and axial-vector couplings to the asymmetry $A_{PV}$. While, we have compared our results for $a_2(x)$ with the presently available results
   in literature, the study of nuclear medium effects in $a_3(x)$ are presented for the first time.
   
   The nuclear medium effects arise due to nuclear structure effects, pionic degrees of freedom, target mass
correction and sea quarks. These effects are calculated in a nuclear model which uses the local density approximation to describe the finite nuclei and a relativistic 
spectral function to describe the momentum distribution of the nucleon. The spectral function is obtained using relativistic
 nucleon propagator in Fermi sea and takes into account Fermi motion, binding energy and nucleon correlations. The mesonic contributions due to pion and rho mesons have been calculated in a 
field theoretical model. The quark and antiquark quarks PDFs have been taken from CTEQ6.6~\cite{cteq} for nucleon and Ref.~\cite{Gluck:1991ey} for mesons. 

We conclude that:  

The nuclear medium effects on $a_2(x)$ are dominated by the nonisoscalarity of the nucleus and lead to an increase in $a_2(x)$ over the entire region of x as compared to the free nucleon 
value and the Cahn-Gilman limit of $\frac{9}{5}-4sin^2\theta_W(=0.90927)$. The enhancement is smaller at lower x and becomes larger at higher x and it is mainly due to the 
nuclear structure effects. In the smaller region of x($x~<~0.4$), there 
is appreciable enhancement due to sea quark effects which is further increased by the nuclear medium effects in the case of nonisoscalar nuclei. The mesonic contribution leads to 
a small decrease at lower values of x($<0.5$) which is almost negligible. This enhancement is 
found to be smaller than the enhancement obtained in the work of Cloet et al.~\cite{thomas}. 

In the case of $a_3(x)$, the nuclear medium effects are found to give a suppression at lower x (x$<$0.5) and an enhancement at higher x (x$>$0.5). 
However, the change in $a_3(x)$ over its free nucleon value is appreciable as compared to the Cahn-Gilman limit of $\frac{9}{5}(1-4sin^2\theta_W)(=0.19656)$
 at low values of x and is mainly due to sea quark content of the nucleon. The quantitative change in $a_3(x)$ due to nuclear medium effects
 is significant for $x$ $<~0.5$. 
 However, the relative contribution of $a_3(x)$ to $A_{PV}$ as compared to $a_2(x)$ is not much affected by the nuclear medium effects,
specially in the region of $x~>~0.4$. Therefore,  any experimental determination of $a_3(x)$ from asymmetry measurements in future will 
have negligible systematic errors due to nuclear medium effects.

Finally the parity violating asymmetry $A_{PV}$ in the scattering of polarised electrons from nuclear targets
is found to increase over its free nucleon value due to nuclear medium effects over the entire region of x. This  enhancement increases with the
nonisoscalarity of the nucleus and is almost negligible for the isoscalar nuclei. The effect of target mass correction is to further
increase the asymmetry by $1-1.5\%$ in the region of $0.3~<~x~<~0.8$. 
These results will be useful in analyzing the experimental data on nuclei 
like $^{12}C$, $^{56}Fe$ and $^{208}Pb$ whenever they become available in future and would help to understand the possible signals for physics beyond the standard model.
\section{Acknowledgments}
M. S. A. is thankful to Department of Science and Technology(DST), Government of India for providing financial assistance under Grant No. SR/S2/HEP-18/2012. 
I. R. S. thanks FIS2011-24149 Spanish project for financial support.
%\end{abstract}

\end{document}